\documentstyle[preprint,aps]{revtex}

\newcommand{\ket}[1]{| #1 \rangle}

\input{psfig}

\begin{document}
\draft
\preprint{5codtol06~~~8/21/96}
\title{Fault-Tolerant Error Correction with Efficient Quantum Codes}

\author{David~P.~DiVincenzo$^1$ and Peter~W.~Shor$^2$}

\address{$^1$IBM T. J. Watson Research Center, Yorktown Heights, NY 10598}
\address{$^2$AT\&T Research, Murray Hill, NJ 07974}

\date{\today}
\maketitle
\begin{abstract}
We exhibit a simple, systematic procedure for detecting and correcting
errors using any of the recently reported quantum error-correcting
codes.  The procedure is shown explicitly for a code in which one
qubit is mapped into five.  The quantum networks obtained are fault
tolerant, that is, they can function successfully even if errors occur
during the error correction.  Our construction is derived using a
recently introduced group-theoretic framework for unifying all known
quantum codes.
\end{abstract}

\narrowtext

The past year has witnessed an astonishing rate of progress in the
development of error-correction schemes for quantum memory and quantum
computation.  The initial discovery\cite{Shor} that a qubit, when
suitably encoded in a block of qubits, can withstand a substantial
degree of interaction with the environment without degradation of its
quantum state, has been followed by myriad contributions which have
identified many new coding schemes
\cite{CS,Steane2,LMPZ,EM,VGW,SS,Lloyd,Schumacher1,KL,Gottes,Steane3,ZL},
considered their application in proposed experimental implementations
of quantum computation\cite{Braun,PVK,Pel}, and established the
relationship of quantum error-correcting codes to the preservation of
quantum entanglement in a noisy environment\cite{purify}.  The most
recent work has unified all the known quantum codes within a
group-theoretic framework\cite{CRSS}.

Throughout the developments of the past year, there has been a hope
that these quantum error-correcting codes would permit quantum
computation to be done fault tolerantly.  Such an outcome was not
guaranteed; in classical computation, the existence of
error-correction codes does not by itself ensure that logic can be
performed using noisy gates.  However, one of us has recently
established a complete protocol for performing fault-tolerant quantum
computation\cite{Shorft}.  The protocol guarantees that, if the loss
of fidelity of the quantum state between the operation of one quantum
gate and the next, due to both decoherence and inaccuracy in the
quantum-gate operation, is $p$, then the number of steps of quantum
computation which can be completed successfully is $O(p^a\exp(b/p^c))$
(for some positive constants $a$, $b$ and $c$), a scaling law which
appears very favorable for the ultimate physical implementation of
large-scale quantum computation.

This fault-tolerant protocol lays down specific rules for how to use
the previously discovered quantum error-correction codes.  The class
of codes first discovered by Calderbank and Shor\cite{CS} and
Steane\cite{Steane2} conform to these rules, and can be used fault
tolerantly; however, it has not been clear that the more efficient
quantum codes which have been discovered more recently (see, e.g.,
\cite{CRSS}) could be utilized in a fault-tolerant computation.  In
this note we establish that errors in all known quantum
error-correcting codes can be corrected in the necessary
fault-tolerant way.  We first show explicitly how this is done in one
of the simplest efficient quantum codes, one which encodes a single
qubit into five\cite{LMPZ,purify}.  This result gives some interesting
insights into the relationship between the different presentations of
this code which have recently appeared in the literature, and it shows
that it is actually necessary to use these different presentations to
produce the fault-tolerant implementation of the error-correction
procedure.  We then show, using the recently developed group-theoretic
framework for the quantum codes, that the protocol developed for the
five-bit code can be generalized to permit all known codes to be used
for error correction in a fault-tolerant way.

We begin with a short review of the five-qubit error-correcting code
as presented in \cite{purify}.  Using this code, an arbitrary qubit
$\ket{\xi}=\alpha\ket{0}+\beta\ket{1}$ is represented by the
five-qubit state $\ket{\xi}=\alpha\ket{c_0}+\beta\ket{c_1}$, where one
choice of the ``code words'' is the pair of basis states
\begin{eqnarray}
\ket{c_0} &=& \ket{00000}\label{sym0}\\
&+& \ket{11000} +\ket{01100} +\ket{00110} +\ket{00011} +\ket{10001}
\nonumber \\
&-& \ket{10100} -\ket{01010} -\ket{00101} -\ket{10010} -\ket{01001}
\nonumber \\
&-& \ket{11110} -\ket{01111} -\ket{10111} -\ket{11011} -\ket{11101}
\nonumber
\end{eqnarray}
and
\begin{eqnarray}
\ket{c_1} &=& \ket{11111}\label{sym1}\\
&+& \ket{00111} +\ket{10011} +\ket{11001} +\ket{11100} +\ket{01110}
\nonumber \\
&-& \ket{01011} -\ket{10101} -\ket{11010} -\ket{01101} -\ket{10110}
\nonumber \\
&-& \ket{00001} -\ket{10000} -\ket{01000} -\ket{00100} -\ket{00010}
\nonumber .
\end{eqnarray}
When encoded in this way, the qubit can survive an interaction with
the environment suffered by any one of the five qubits.  For purposes
of error correction, it is sufficient to take the error caused by the
environment to be of three different types\cite{EM,purify}: bit $i$
may suffer a bit-flip error, represented by the operator $X_i$ acting
on coded state $\ket{\xi}$; it may suffer a conditional phase-shift
error ($Z_i$), or it may suffer both simultaneously ($Y_i$).  (We use
the notation of Refs.~\cite{Gottes,CRSS}.)  The right-hand column of
Table~\ref{table1} lists the 16 possible error processes $P$
(including the no-error case $P=I$).  During error correction, the
erroneous state $P\ket{\xi}$ is subjected to some quantum-computation
operations (one- and two-bit quantum gates\cite{G9}) so that
measurements on some of the qubits will reveal the identity of the
error process $P$, without disturbing the superposition of code
words.  When the error process is determined, the effect of $P$ can be
undone, returning the qubit to its undisturbed state $\ket{\xi}$.

\begin{figure}

\vspace{1in}

\centerline{\psfig{figure=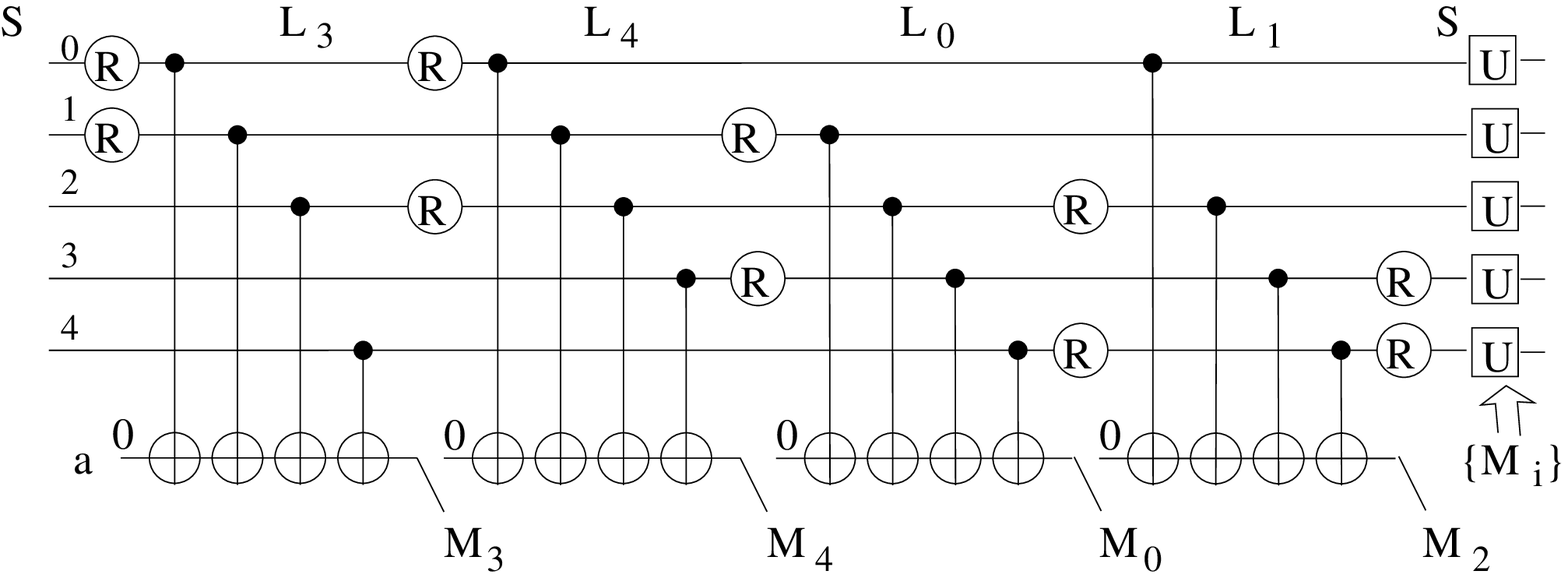,width=8.0in}}

\vspace{1in}

\caption{Quantum network to correct for one-bit errors in the 5-bit
code in the $S$ presentation.  Four different code presentations
$L_{3,4,0,1}$ are used in the different stages of error detection.  By
a simple modification of the ancilla space $a$, and by appropriate
repetitions of the syndrome computation, this error-correction network
can be made fault tolerant.
\label{5bell}}
\end{figure}

It has now been shown by a number of authors\cite{LMPZ,purify,Braun}
that there exist various quantum circuits which perform the necessary
error correction on the five-bit coded state.  However, none of them
perform this error correction fault tolerantly (unlike the network of
Fig.~\ref{5bell} which can operate fault tolerantly).  We call a
quantum error-correcting network fault tolerant if it can recover from
errors {\it during} the operation of the network.  Previous
constructions are not fault tolerant because they use two-bit quantum
gates involving pairs of qubits within the coded state.  If an error
occurs on one of these qubits before or during the operation of this
two-bit gate, the error will, in general, propagate to both of the
qubits, and to yet others if additional two-bit operations are
performed.  In the five-bit code, two errors are already more than can
be recovered from, so such two-bit gates must be avoided.  The network
of Fig.~\ref{5bell} avoids them by using only two-bit gates which
connect the coded bits to ancilla bits $a$, so that, with small
modifications, it can be made perfectly fault tolerant.  These
modifications are described briefly in \cite{Shorft} and given in
detail in \cite{PVKft}.

To explain how the network of Fig.~\ref{5bell} works, we note that the
code of Eqs.~(\ref{sym0},~\ref{sym1}) can be presented in an infinite
number of ways, all related by a change of basis of any one of the
five qubits.  Even if we confine ourselves to bases in which the
superpositions all involve equal amplitudes as in
Eqs.~(\ref{sym0},~\ref{sym1}), the number of alternative presentations
is very large.  One important class of presentations is symmetric
under cyclic permutation of the five qubits, as in the example given
above.  We will define a particular symmetric presentation, $S$, as
the one in which $\ket{0}$ is coded as $\ket{c_0}+\ket{c_1}$, and
$\ket{1}$ is coded as $\ket{c_0}-\ket{c_1}$.

Another class of presentation has been given in the work of Laflamme
{\it et al.}\cite{LMPZ}.  Their presentation is obtained by starting
with presentation $S$ and applying the one-bit rotation $R =
\frac{1}{\sqrt{2}}
\renewcommand{\arraystretch}{.33}
\left(\begin{array}{cc} \scriptstyle 1& \scriptstyle ~1 \\ \scriptstyle
1 & \scriptstyle -1\end{array}\right)$
\renewcommand{\arraystretch}{1}
to qubits 0 and 1 (we number the qubits 0--4 as in Fig.~\ref{5bell}).
In this presentation, the code words are
\begin{eqnarray}
\ket{c'_0} &=&
    \ket{00010} +\ket{00101} -\ket{01011} +\ket{01100} \label{laf0} \\
&+& \ket{10001} -\ket{10110} -\ket{11000} -\ket{11111},\nonumber
\end{eqnarray}
and
\begin{eqnarray}
\ket{c'_1} &=&
    \ket{00000} -\ket{00111} +\ket{01001} +\ket{01110} \label{laf1} \\
&+& \ket{10011} +\ket{10100} +\ket{11010} -\ket{11101}.\nonumber
\end{eqnarray}
We will call this presentation $L_3$; except for a trivial relabeling
of the qubits, this is exactly the one given in \cite{LMPZ}.  The
reason for the subscript is that, since the $L_3$ presentation is {\it
not} symmetric under cyclic permutation, there are five distinct ones
$L_{0-4}$.  The particular label 3 is used for this example because of
an important property which this presentation possesses: all the basis
states of both the code words in Eqs.~(\ref{laf0}, \ref{laf1}) have
even parity for the group of four qubits 0, 1, 2, and 4.  Thus, a
convenient label for this presentation is the qubit which is left out
of this parity.  Since an error can change this parity, we can learn
one bit of information about the error process by collecting up this
parity into the ancilla qubit $a$ (done by the first four quantum XOR
gates in Fig.~\ref{5bell}), and performing measurement $M_3$ on $a$.

The remainder of the quantum circuit in Fig.~\ref{5bell} is
self-explanatory.  By passing in succession into three additional
bases, those corresponding to the code presentations $L_4$, $L_0$, and
$L_1$, three additional parity bits may be obtained in measurements
$M_4$, $M_0$, and $M_1$.  (In standard coding theory terminology, the
outcome of these four measurements is called the {\it error
syndrome.})  As Table~\ref{table1} indicates, these measurements
uniquely distinguish the error process~$P$.  This error can then be
undone by returning the code to the original $S$ basis and selecting
the appropriate one-bit operation $U$.

As presented, this error-correction network is not completely fault
tolerant, because an error occurring on one of the $a$ bits can be
transmitted back to one of the code qubits through the action of the
XOR gates.  For instance, if a phase error occurs on the ancilla qubit
$a$ between the second and third XOR gates in Fig.~\ref{5bell}, the
back action of the XOR gates results in two phase errors in the state
of the code qubits, rendering them uncorrectable.  However, as one of
us has recently shown\cite{Shorft}, the network may be made completely
fault tolerant by replacing the single-bit ancilla $a$ by a set of
four qubits, each of which is initialized to a ``cat'' state
$\ket{0000}+\ket{1111}$.  If the targets of each the XOR gates are
four different qubits in the cat state, then the parity of the
measured state of the four ancilla bits gives the same information as
the measurements indicated in Fig.~\ref{5bell}.  However, the
back-action that makes the errors on the ancilla $a$ dangerous is
avoided.  The ancilla errors may still result in a mistake in the
measured syndrome; we prevent this from adding errors to the coded
state by repetition of the entire network and syndrome measurement,
before the one-bit operation $U$ is performed\cite{Shorft}.  Once the
correct syndrome has been confirmed, the correct $U$ may be
applied\cite{PVKft}.

The fact that the four measurements $M_{3,4,0,1}$ completely
distinguish the error process is no accident; it is guaranteed by the
group-theoretic structure of these codes\cite{CRSS,Gottes}.  In fact,
the procedure devised above can be generalized to give a
fault-tolerant error-correction procedure that covers every quantum
code which is presently known, all of which are derivable as
eigenspaces of Abelian subgroups of a group $E$\cite{Kerdock},

The group $E$ is obtained by taking all products of the $X_i$, $Y_i$
and $Z_i$ operators introduced above.  Given an Abelian subgroup $G$
of $E$ containing $2^g$ elements, the matrices representing $G$ can be
simultaneously diagonalized (because they commute with each other).
This yields $2^g$ eigenspaces each of dimension $2^{n-g}$.  Choosing
any of these eigenspaces gives a quantum code mapping $n-g$ qubits
into $n$ qubits, and the error correction properties of this code can
be derived from the combinatorial properties of the subgroup $G$
\cite{Gottes,CRSS}.  The subgroup $G$ can be generated by an
independent set of $g$ of its elements, which we call generators;
again, these generators are products of the $X_i$, $Y_i$, and $Z_i$
operators.  For instance, one of the generators for the five-bit code
in the $S$ presentation is, in the notation of \cite{CRSS},
$X(11000)Z(00101)$; a 1 in the $i^{th}$ place in the $X$ list means
that $X_i$ is included in the operation, a 1 in the $Z$ list means
that $Z_i$ is included, and a 1 in both lists means that $Y_i$ is
included.

Each such generator of $G$ gives a prescription for one stage of
fault-tolerant error correction, as follows: First, a change of basis
involving just one-bit operations is performed, in order to place the
generator in the form $X(000...0)Z(z_1z_2z_3...z_n)$  where
$z_i=0$ or $1$ (i.e., so that the generator contains only
$Z_i$ factors).  The one-bit rotation required for the $i^{th}$ qubit
is easily determined: if $X_i=0$ do nothing, if $X_i=1$ and $Z_i=0$,
apply $R$ to the $i^{th}$ qubit, and if $X_i=Z_i=1$, apply $R'$, where
$R' = \frac{1}{\sqrt{2}}
\renewcommand{\arraystretch}{.33}
\left(\begin{array}{cc} \scriptstyle 1& \scriptstyle i \\ \scriptstyle
i & \scriptstyle 1\end{array}\right)$.
\renewcommand{\arraystretch}{1}
After this change of basis, the
non-zero elements of the new $Z$ bit string will be just those for which
$X$ or $Z$ were non-zero in the original basis.  The next step of the
error correction is to collect up and measure the parity of the bits
with non-zero entries in the $Z$ string, using the ancilla technique
discussed above.
Finally, undo the basis transformation.  Repeat this
procedure for each generator of $G$.

It is guaranteed that this set of measurements will completely
determine the error process~$P$.  The measurement on a quantum state
corresponding to one of the generator matrices of $G$ gives the
eigenvalue of the quantum state with respect to that matrix, reducing
the number of eigenspaces which the quantum state might lie in by a
factor of 2.  Thus, if the measurements are made for every matrix in a
generator set for the subgroup $G$, this guarantees that the complete
set of eigenvalues for this state with respect to the subgroup is
known.  This complete set of eigenvalues places the quantum state
uniquely in one of the eigenspaces.  The error processes $X_i$, $Y_i$
and $Z_i$ permute these eigenspaces\cite{CRSS}, so knowing which
eigenspace a state belongs to is enough to uniquely determine the
unitary transformation $U$ of Fig.~\ref{5bell} which will correct the
error.  ($U$ is also one of the unitary transformations $X_i$, $Y_i$
or $Z_i$.)  The requirement that all the measurements be
simultaneously observable can be seen to be the physical justification
for the requirement that all the generator matrices commute.

The number of gates this construction gives for error correction of a
quantum code can be estimated.  Suppose it is applied to a quantum
code mapping $k$ qubits into $n$ qubits, correcting $t$ errors.  (Many
such codes have now been tabulated~\cite{Steane3,CRSS}.)  The syndrome
will contain $n-k$ bits, and computing each bit of this syndrome
requires at most $n$ XOR gates.  Similarly, between $0$ and $n$
rotation gates will also be required before and after the computation
of each of the bits of the syndrome.  Thus, the number of gates
required by this technique for an $n$-qubit code is at most
$2n(n-k+1)$, and the number of ancilla bits needed is no greater than
$n(n-k)$.  The suitable use of this error-correction network will
be fault tolerant: up to $t$ errors can occur during the error
correction process itself without irretrievably damaging the state of
the $k$ coded qubits.

The class of quantum error-correcting codes given in \cite{CS,Steane2}
have generators which are either products only of $Z$'s or only of
$X$'s.  This technique applied to these codes thus reduces to first
finding the parity of sets of qubits corresponding to the generators
composed of $Z$'s, next applying the basis transformation $R$ to each
qubit, then finding the parities corresponding to generators composed
of $X$'s, and finally undoing the basis transformation $R$ on each
qubit.  This is exactly the prescription given by Steane
\cite{Steane2}.  For this class of codes, the correction procedure for
bit-flip ($X$) errors can be decoupled from the treatment of phase
($Z$) errors.  The bit-flip ($X$) errors affect the eigenvalues of
matrices which are a product of $Z$'s, and vice versa.  Each type of
error can be thought of classically (in the appropriate basis) and
corrected using classical techniques, as is emphasized in Steane
\cite{Steane2}.

To conclude, we have shown that the group-theoretic structure of all
the reported quantum error-correcting codes provides rules for
designing very simple quantum networks to detect errors and restore
the quantum system to its undisturbed state.  These networks are
superior to previously reported ones in that they can be implemented
in a fault-tolerant way.  We note that our result does not provide a
complete solution for how to use the most efficient quantum codes in
fault-tolerant quantum computation, since this would require a
fault-tolerant implementation of multi-bit gates on the coded
qubits\cite{Shorft}.  Such fault-tolerant gate implementations are
known for the non-optimal codes of \cite{CS,Steane2}, but it is not
yet clear that they exist for all the codes derived from the group $E$
(however, see \cite{ZL}).  Even without this, though, it is clear that
the procedures developed here may ultimately have a variety of
applications for quantum memory, quantum communications, and quantum
computation.

We would like to thank Rob Calderbank for helpful discussions.

\begin{table}
\caption{The four measurement outcomes in the fault-tolerant error
correction, and the error process $P$ revealed by each.
\label{table1}}
\begin{tabular}{cccccc|cc}
&$M_3$ & $M_4$ & $M_0$ & $M_1$ && {\it P} &\\
\tableline
& 0 & 0 & 0 & 0 && {\em I} &\\
& 0 & 0 & 0 & 1 && $Z_4$ &\\
& 0 & 0 & 1 & 0 && $X_1$ &\\
& 0 & 0 & 1 & 1 && $Z_3$ &\\
& 0 & 1 & 0 & 0 && $X_3$ &\\
& 0 & 1 & 0 & 1 && $X_0$ &\\
& 0 & 1 & 1 & 0 && $Z_2$ &\\
& 0 & 1 & 1 & 1 && $Y_3$ &\\
& 1 & 0 & 0 & 0 && $Z_0$ &\\
& 1 & 0 & 0 & 1 && $X_2$ &\\
& 1 & 0 & 1 & 0 && $X_4$ &\\
& 1 & 0 & 1 & 1 && $Y_4$ &\\
& 1 & 1 & 0 & 0 && $Z_1$ &\\
& 1 & 1 & 0 & 1 && $Y_0$ &\\
& 1 & 1 & 1 & 0 && $Y_1$ &\\
& 1 & 1 & 1 & 1 && $Y_2$ &\\
\end{tabular}
\end{table}

\end{document}